# Cultural Prompting Improves the Empathy and Cultural Responsiveness of GPT-Generated Therapy Responses

Serena Jinchen Xie, MS[1], Shumenghui Zhai, RN, MPH, PhD[2], Yanjing Liang, MS[1], Jingyi Li, RN, PhD[3], Xuehong Fan, MS[1], Trevor Cohen, MBChB, PhD[1], Weichao Yuwen, RN, PhD[3]

[1]University of Washington, Seattle, WA; [2]Pacific Lutheran University, Tacoma, WA, USA; [3]University of Washington, Tacoma, WA

**Abstract**

*Large Language Model (LLM)-based conversational agents offer promising solutions for mental health support, but lack cultural responsiveness for diverse populations. This study evaluated the effectiveness of cultural prompting in improving cultural responsiveness and perceived empathy of LLM-generated therapeutic responses for Chinese American family caregivers. Using a randomized controlled experiment, we compared GPT-4o and Deepseek-V3 responses with and without cultural prompting. Thirty-six participants evaluated input-response pairs on cultural responsiveness (competence and relevance) and perceived empathy. Results showed that cultural prompting significantly enhanced GPT-4o's performance across all dimensions, with GPT-4o with cultural prompting being the most preferred, while improvements in DeepSeek-V3 responses were not significant. Mediation analysis revealed that cultural prompting improved empathy through improving cultural responsiveness. This study demonstrated that prompt-based techniques can effectively enhance the cultural responsiveness of LLM-generated therapeutic responses, highlighting the importance of cultural responsiveness in delivering empathetic AI-based therapeutic interventions to culturally and linguistically diverse populations.*

**Introduction**

Large language model (LLM)-based conversational agents (chatbots) have emerged as promising tools to bridge mental health service gaps by providing cost-effective and on-demand therapeutic support,[1] offering immediate emotional support, stress management strategies, and culturally relevant coping techniques[2]. However, research shows that many LLMs face critical challenges in cultural alignment, often misinterpreting culturally specific expressions or providing advice that does not align with the traditions, beliefs, and family dynamics of diverse communities[3]. In particular, studies have demonstrated that LLM-generated responses tend to align more closely with Western cultural values than those from non-Western backgrounds, which can result in responses that feel culturally insensitive or irrelevant to users from minority communities[4], making it difficult to be effectively used in mental health care for diverse populations[5]. There is a critical need for culturally responsive LLM-based mental health interventions.

Culturally responsive therapy is the approach that considers the interaction between a client's presenting problem and their culture[6]. Cultural responsiveness goes beyond the cultural sensitivity of therapists. It consists of understanding and respecting cultural differences and actively adapting and tailoring one's approach to ensure relevancy, acceptability, and effectiveness for the client's cultural context[7]. The importance of cultural responsiveness in mental health care has been emphasized for decades[8], with evidence indicating that cultural insensitivity in therapy can lead to a lack of trust between patients and providers and alliance ruptures[9]. Empathy, a core mechanism of building therapeutic alliances and subsequent changes, requires cultural contextualization to be effectively conveyed in clinical interactions. Elliott et al.[10] highlighted that empathy strongly predicted positive therapy outcomes. However, without cultural responsiveness, any therapeutic interaction may fail to recognize a client's cultural background, and empathy can be misunderstood or perceived as a superficial approach, leading to reduced engagement and less effective treatment[11]. We believe that cultural responsiveness has the potential to deepen empathy by making therapeutic interactions more contextually relevant and resonant with clients, deepening the sense that clients feel heard and supported.

Cultural adaptation of psychosocial interventions originated with the application of the ecological validity theory[12] to adapt interventions to specific cultural patterns, meanings, and values. Traditional adaptation methods involve an iterative and collaborative process that includes input from individuals within the target populations[3,13]. However, this presents challenges for AI-driven mental health interventions. Unlike conventional interventions with static and manually written content, LLM-driven chatbots generate dynamic responses in real-time, rendering traditional expert-guided cultural adaptation approaches inapplicable[13]. To date, there is a gap in the literature on methodologies for culturally adapting LLM-driven mental health chatbots. Given the dynamic nature of LLM responses, new methodologies are required to ensure that AI-generated content is culturally responsive.

Previous research has shown that merely prompting in the language of the intended culture is not sufficient for cultural adaptation[14]. Subsequent approaches have attempted to modify model behavior using more explicit methods. There are two general directions of model adaptation: prompt-based methods (e.g., prompt tailoring) and weight-changing methods (e.g., fine-tuning). For example, Soaiman and Dennison fine-tuned LLMs on a value-targeted dataset[15], and Nguyen et al.[16] continued pre-trained LLMs to capture more Southeast Asia cultural norms, customs, and stylistic preferences[16]. Non-weight-changing methods rely on prompts to modify model behavior and require fewer resources compared to weight-changing approaches. One example is the cultural prompting technique, which incorporates cultural contexts as part of prompts (e.g., "You are an average human being born in [country/territory] and living in [country/territory] responding to the following survey question.")[5]. Tao et al.[5] demonstrated that cultural prompting can effectively reduce cultural bias in GPT. In another work that quantified the cultural alignment of LLMs, GPT-4 showed the capability to adapt to cultural nuances in response to an explicit cultural context prompt, particularly in Chinese settings[17]. However, other research yielded mixed results on the effectiveness of cultural prompting[18]. Beck et al.[19] questioned the readiness of prompt-based methods by highlighting the lack of robustness, being highly sensitive to prompt formulation, model choice, and specific tasks performed. Mukherjee et al.[20] observed that model responses varied significantly for both culturally sensitive and non-sensitive prompts, suggesting that cultural conditioning may be a 'placebo effect' rather than actual cultural adaptation.

While prior studies have investigated the cultural biases embedded within LLMs—including their alignment with cultural values[14,17], interpretations of cultural proverbs[21], and knowledge of cultural norms[22–24]—these works have primarily focused on probing model characteristics rather than evaluating their effectiveness in real-world therapeutic interactions. The most commonly used tools for probing cultural alignment were Hofstede's Value Survey Model (VSM)[25] and World Value Survey (WVS)[26] for quantifiable evaluation. VSM and WVS are two influential frameworks for cross-cultural analysis, where VSM categorizes national cultures in six dimensions and WCS is a global research that maps cultural values through representative population sampling. Studies then probed the cultural alignment of LLMs by reformulating the survey questions from WVS and VSM into cloze-style (fill-in-the-blank) questions, analyzing the models' predicted word probabilities, and comparing the predictions with established cultural value survey results[14,17]. A recent survey provided a comprehensive overview of cultural alignment methodologies, benchmarks, and evaluation approaches[27]. However, for healthcare, cultural responsiveness should be situated within the context of treatment delivery[12]. Yet, to our knowledge, no study has systematically evaluated how an AI-driven mental health chatbot's cultural responsiveness affects user experiences in treatment settings or with human-centered evaluation approaches. As Blodgett et al.[28] highlighted, NLP models are ultimately integrated into real-world applications that directly impact individuals' lives. In the context of mental health care, evaluation methods should adopt a human-centered perspective, prioritizing users' needs, values, and interaction behaviors to ensure AI-driven interventions have meaningful and desirable real-world effects.

The current work addresses these gaps in the literature by improving and evaluating the cultural responsiveness of LLM-generated responses in a therapy-delivery context. To this end, we used a prompt-based technique, cultural prompting, to adapt the responses of an AI therapist for Chinese American family caregivers. Asian Americans are the fastest-growing racial and ethnic minority group in the United States, with approximately 67% of Asian adults being immigrants[29]. Among Asian American subgroups, Chinese Americans represent the largest part, making up

24% of the Asian American population[30]. Studies indicate that Chinese American caregivers experience higher levels of psychological distress but remain underserved by mental health services due to reasons such as a lack of linguistically appropriate services[31]. Our objectives were to conduct a comparative assessment of culturally adapted versus baseline LLM-generated interactions and to assess 1) how effective cultural prompting is in improving the cultural responsiveness of AI therapist responses and 2) whether improving cultural responsiveness of these responses also improves perceived empathy. Our study aims to evaluate the hypothesis that a culturally responsive AI therapist will also be perceived as more empathetic, which could lead to better clinical outcomes (Figure 1). This research addresses a critical gap by introducing and evaluating LLM-based methodologies for real-time cultural adaptation in AI-driven mental health applications. By promoting the cultural responsiveness of an AI therapist, we seek to make mental health care more inclusive, engaging, and effective. Findings from this study will help advance the development of culturally responsive AI-driven mental health interventions and improve their accessibility and effectiveness for diverse caregiving populations.

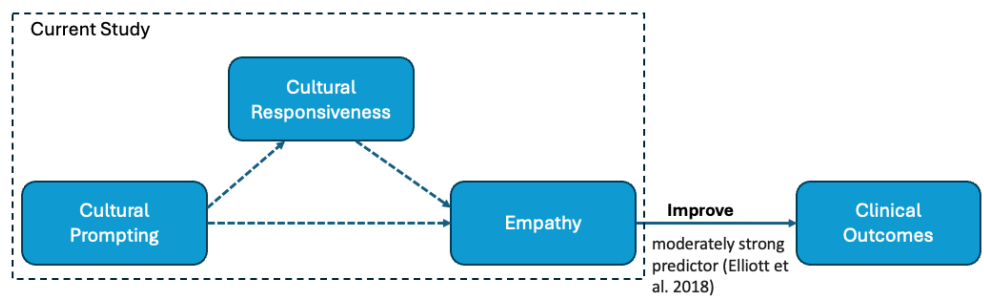

**Figure 1.** Illustration of the hypothesis tested in the current study and the relationship with clinical outcomes.

**Methods**
We conducted a randomized controlled experiment in the following steps: 1) curating high-quality, culturally relevant example user inputs, 2) generating LLM responses using two LLMs with and without cultural prompting techniques (2x2 design), and 3) a blinded, randomized mixed-methods assessment of input-response pairs across four model configurations.

Cultural Representative Input Development and Response Generation
We curated a set of example family caregiver inputs by working with six community partners who have rich experiences working with Chinese American family caregivers and are familiar with family caregivers' challenges. Through weekly meetings, we developed six personas of Chinese caregivers. Each community partner generated four example client inputs in simplified or traditional Chinese per persona, following detailed guidelines. In total, we collected 144 examples of caregiver inputs in Chinese. After that, the research team selected 45 inputs with rich cultural contexts for response generation. Inputs were excluded if too general (e.g., "I am too stressed"), not culturally salient (e.g., "Every time I see the hospital bills, I'm getting a headache."), or similar to an already selected input.

We utilized GPT-4o and Deepseek-V3 to generate responses to the selected caregiver inputs, using system prompts tested in a previous study[32]. Our model selection was motivated by variations in model characteristics (e.g., open-source vs. closed-source), performance metrics, and potential cultural competencies. GPT-4o, a closed-source model developed by OpenAI, represents the current benchmark in general language capabilities[33] and has demonstrated robust performance across multilingual contexts[34]. For comparison, we selected Deepseek-V3 as the open-source model. We chose Deepseek-V3 because it is trained on the multilingual corpus, predominantly consisting of English and Chinese text. Deepseek-V3 outperforms other open-source models, such as LLaMA-3, across most benchmarks while achieving comparable performance to leading closed-source models like GPT-4o and Claude-Sonnet-3.5. Notably, Deepseek-V3 exceeds both GPT-4o and Claude-Sonnet-3.5 in Chinese factual knowledge benchmarks, particularly Chinese SimpleQA[35]. This deliberate comparison of models enables examination of whether models with different training paradigms yield varying degrees of cultural responsiveness with and without explicit cultural prompting. Moreover, previous research has demonstrated that using English prompts could homogenize cultural differences in the responses from LLMs[18], suggesting the importance of constructing prompts directly in the target language. Therefore, all system prompts were translated into Chinese by the authors. For each model, we used two prompt versions: with and without cultural prompting. The prompts with cultural prompting included the following sentence (translated into Chinese) as the first sentence in the system prompt: "When responding to caregivers, please fully consider their Chinese cultural background. Use language, values, and communication styles that align with Chinese culture to ensure your message resonates with them." The rest of the prompt was the same for both versions. This design yielded four distinct response versions (Table 1) for each user input, allowing for a systematic comparison of cultural competency across prompting techniques and models.

Table 1. Four model configurations using two different LLMs with and without cultural prompting.

| Model id | LLM | With Cultural Prompting Techniques? |
| --- | --- | --- |
| 1 | GPT-4o | No |
| 2 | GPT-4o | Yes |
| 3 | Deepseek-V3 | No |
| 4 | Deepseek-V3 | Yes |

Recruitment and Data Collection

We employed a combination of snowball sampling and targeted outreach to recruit participants. We distributed study flyers via social media platforms and recruited participants at two adult daycare centers primarily serving Chinese Americans. All evaluations were conducted through the Qualtrics online surveys. Each participant was instructed to assess four sets of five input-response pairs, randomly selected and with presentation order randomized to mitigate sequence effects. Participants evaluated each set on cultural responsiveness and perceived empathy. We used two complementary measures to assess cultural responsiveness: 1) the Healthcare Provider Cultural Competency Measure (CCM)[36], is a validated patient-reported instrument that captures patients' perceptions of their providers' cultural competency. This 9-item evaluation evaluates three critical domains—cultural knowledge, awareness, and skill. This scale has established reliability (Cronbach's alpha 0.68–0.98) in previous studies[37–39]. To minimize participant burden while maintaining construct validity, we selected a representative item from each domain plus one overall assessment item, preserving the theoretical structure while optimizing completion time. The internal consistency reliability of the four CCM items was assessed using Cronbach's alpha[40] ($\alpha = 0.96$), indicating an excellent internal consistency for the selected CCM items and supporting the validity of this shortened form. 2) The Cultural Relevance Questionnaire (CRQ) focused on cultural sensitivity[55] and ecological validity theory[42]. The CRQ has demonstrated utility in assessing internet-delivered therapeutic interventions[43] and low-intensity psychological approaches[44], making it particularly appropriate for our low-intensity therapy intervention context. Prior studies report good internal consistency (Cronbach's a > 0.70)[44,45]. The Chrobach's alpha for CRQ in our study is 0.94. By employing both CCM and CRQ, we aimed to probe cultural responsiveness comprehensively. Empathy was measured by a 3-item instrument specifically used to evaluate the empathy expressed in text-based mental health support[46]. Specifically, the instrument assesses three aspects of empathy: emotional reactions (expressing emotions such as warmth, compassion, and concern), interpretations (communicating an understanding of feelings and experiences inferred from the user input), and explorations (exploring the feelings and experiences not stated in the user input).

Following quantitative ratings, participants selected their most and least preferred response versions and provided rationales through an open-ended prompt: "Can you share why you like/dislike the responses?" Demographic data collected included age, gender, years living in the U.S., education level, and self-reported English proficiency level.

Data Analysis

We conducted a comprehensive analysis combining descriptive and statistical methods. For descriptive analysis, we computed means, standard deviations, and ranges for each evaluation measure (cultural competence (CCM), cultural relevance (CRQ), and perceived empathy) across four model configurations. Inter-rater reliability was assessed using intraclass correlation coefficients. We examined preference patterns across the four configurations using Pearson's chi-square test against the null hypothesis of random distribution, followed by a post-hoc analysis to identify specific model configurations contributing statistically significantly to the chi-square value. We analyzed correlations between ratings and participant demographics using Spearman's Rank Correlation for numerical variables and the Mann-Whitney U test for categorical variables. We conducted statistical comparisons using Wilcoxon signed-rank tests to compare ratings between 1) responses generated by different LLMs and 2) responses generated with and without cultural prompting. To test the hypothesis that perceived empathy is primarily mediated by cultural relevance rather than direct cultural prompting, we conducted an analysis of covariance (ANCOVA) test, examining empathy rating differences between GPT-4o's responses with and without cultural prompting, controlling for differences in cultural relevance ratings by using differences in CRQ and CCM ratings as covariates. Statistical significance was considered established at $p<0.05$.

**Results**

In total, 36 participants participated in the study. Participants were predominantly female (n=28, 77.8%) with a mean age of 50.8 years (SD=9.8, range=28-70). The sample represented a well-established immigrant population with an average U.S. residency duration of 20.9 years (SD=8.0, range=8-35). Education levels were notably high, with the majority holding graduate or professional degrees (n=22, 61.1%), followed by Bachelor's degrees (n=10, 27.8%), some college education (n=3, 8.3%), and one participant declining to specify (2.8%). Regarding English language proficiency, most participants self-reported high proficiency levels (n=19, 52.8%), with the rest indicating moderate (n=12, 33.3%) or limited proficiency (n=5, 13.8%).

Cultural responsiveness and empathy among different models

The descriptive statistics of ratings for the three evaluation measures across different model configurations are shown in Table 2. GPT-4o with cultural prompting was consistently rated the highest across all three assessment dimensions. The inter-rater reliability analysis yielded an intraclass correlation coefficient of 0.86 (95% CI [0.79, 0.91]), indicating good reliability of the ratings across the 36 evaluators.

**Table 2.** Descriptive statistics of the ratings for responses generated by each model configuration.

| Model Configuration | CCM (7-point Likert Scale, mean (SD, range)) | CRQ (5-point Likert Scale, mean (SD, range)) | Empathy (5-point Likert Scale, mean (SD, range)) |
|---|---|---|---|
| GPT-4o without cultural prompting | 4.01 (1.65, 1-6.75) | 2.96 (1.19, 1-5) | 3.52 (1.02, 1-5) |
| GPT-4o with cultural prompting | 5.00 (1.61, 1-7) | 3.63 (1.09, 1-5) | 3.94 (0.94, 1.67-5) |
| Deepseek without cultural prompting | 4.08 (1.72, 1-6.75) | 2.98 (1.22, 1-5) | 3.39 (0.98, 1-5) |
| Deepseek with cultural prompting | 4.26 (1.53, 1-7) | 3.22 (1.00, 1-5) | 3.41 (0.88, 1.67-5) |

The Wilcoxon Signed-Rank test showed that cultural prompting significantly enhanced performance across all measured dimensions for GPT-4o with relatively large effect sizes ($r>0.5$) but not for Deepseek-V3 (Table 3). With cultural prompting, the GPT model showed statistically significantly better performance compared to Deepseek with cultural prompting across all measured dimensions with large effect sizes ($r>0.5$).

We employed Spearman's Rank Correlation and the Mann-Whitney U test to examine associations between ratings and participant demographics (age, gender, duration of time living in the U.S., education level, self-reported English proficiency level). Analysis revealed no statistically significant correlations between any demographic characteristics and rating patterns across responses generated by different model configurations.

**Table 3.** Wilcoxon Signed-Rank Test Results for Between-Condition Comparisons.

| Comparison | Cultural Competency (CCM) | Cultural Relevance (CRQ) | Empathy |
|---|---|---|---|
| GPT-4o with cultural prompting **vs.** without | W=55.0, p<.001, r=.74* | W=68.5, p<.001, r=.71* | W=91.5, p=.01, r=.66* |
| DeepSeek-V3 with cultural prompting **vs.** without | W=204.0, p=.11, r=.39 | W=177.0, p=.16, r=.46 | W=188.0, p=.98, r=.43 |
| GPT-4o **vs.** DeepSeek-V3 (both without cultural prompting) | W=149.5, p=.09, r=.52 | W=172.5, p=.49, r=.47 | W=139.5, p=.36, r=.54 |
| GPT-4o **vs.** DeepSeek-V3 (both with cultural prompting) | W=117.5, p=.01, r=.60* | W=148.5, p=.02, r=.52* | W=110.5, p=.01, r=.61* |

* indicates statistically significant differences at p<.05.

Model preferences

The chi-square analysis on the relationship between model configuration and participants' preferences revealed a statistically significant association between model configuration and participants' preferences ($\chi^2(3) = 23.78$, $p < .001$). Post-hoc analysis of the residuals indicated that responses generated by GPT-4o with cultural prompting were significantly more preferred than expected under the null hypothesis (randomly distributed preference) with a standardized residual equal to 2.21. Qualitative feedback showed that users felt GPT-4o with cultural prompting provided a balanced mix of emotional resonance, practical recommendations, and encouragement. They noted that culturally tuned responses did a better job of "expressing care, providing concrete suggestions, and offering thoughtful reasoning." On the other hand, responses generated by GPT-4o without cultural prompting demonstrated a trend toward being the least preferred, though this did not reach statistical significance based on the post-hoc analysis. Relevant qualitative feedback includes "too official" and "sounded like ChatGPT." Others described them as "empty, repetitive, and indirect" and expressed frustration that the chatbot "avoided giving direct responses" or was "too many words without getting to the point." Although quantitative ratings and the preference votes did not show Deepseek with cultural prompting was preferred, some participants noted Deepseek with cultural prompting as a "culturally focused and individualized conversation compared to the other groups" and had "a strong sense of empathy and understanding of Chinese culture." On the contrary, responses from Deepseek without cultural prompting were described as "(the responses) not specifically targeting the cultural background of Chinese users." These responses suggested that responses generated by LLMs with cultural prompting enabled a deeper emotional connection and made participants feel more understood. To illustrate the qualitative differences between model configurations, we present two example responses (translated into English) to the same user input. GPT-4o, without cultural prompting, responded: "*This is indeed a very difficult decision, because you want to make the best choice for your mother. It's completely normal to feel uncertain in a situation like this. Could you share what your biggest challenge or concern is when making this decision?*" In contrast, GPT-4o with cultural prompting responded: "*I understand your distress and inner conflict. Your sense of duty and filial piety are clear. No matter your decision, it reflects your care for your mother. Could you share what aspect of caregiving is currently causing you the most stress?*"

**Table 4.** Counts of each model's responses are selected as the most and least preferred models.

| Model Configuration | Most preferred Counts | Least Preferred Counts |
|---|---|---|
| GPT-4o without cultural prompting | 5 | 15 |
| GPT-4o with cultural prompting | 19 | 4 |
| Deepseek without cultural prompting | 4 | 10 |
| Deepseek with cultural prompting | 8 | 7 |

Cultural responsiveness mediates the relationship between cultural prompting and empathy

Results demonstrated the strong explanatory power of adjusted $R^2 = 0.41$ ($F(1,34) = 24.94$, $p<.0001$). The regression coefficient for cultural relevance difference ($\beta = 0.54$, SE = 0.108, t = 4.90, $p < .0001$, 95% CI [0.32, 0.76]) indicates that with a 1 point increase in cultural relevance rating difference, the empathy rating difference increases about 0.54 point in the empathy rating scale. When controlling for cultural relevance as a covariate, the independent effect of adding cultural prompting on empathy rating difference became non-significant ($\beta = 0.05$, p = 0.70). This result provided evidence that differences in perceived empathy between the configurations were predominantly mediated by differences in cultural relevance rather than resulting directly from prompt modifications.

Similarly, we tested the hypothesis that cultural competency mediated the perceived empathy in the GPT-4o responses. A similar mediation pattern emerged for cultural competence as a mediator for the change in empathy. The ANCOVA model demonstrated strong statistical power ($F(1, 34) = 34.94$, $p < .0001$, adjusted $R2=0.49$), with a non-significant intercept term ($\beta = -0.068$, SE = 0.13, t = -0.510, p = .61), indicating that when controlling for cultural competence as a covariate, the independent effect of adding cultural prompting on empathy rating difference became non-significant. The regression coefficient of cultural competence difference ($\beta = 0.49$, SE = 0.083, t = 5.91, $p < .001$, 95% CI [0.32, 0.66]) revealed a strong positive relationship between the improvements in cultural competence and the corresponding improvement in perceived empathy. Specifically, a 1-point increase in cultural competence rating difference corresponds to about 0.49 points in empathy rating difference.

**Discussion**

This study evaluated the cultural responsiveness (relevance and competence) and empathy of AI-generated therapeutic responses for Chinese American family caregivers through a randomized controlled comparison of two LLMs (GPT-4o and Deepseek-V3) with and without a cultural prompting technique. Our findings demonstrate that cultural prompting enhances GPT-generated responses' perceived cultural responsiveness and empathy. Although the two LLMs showed comparable performance without cultural prompting, GPT-4o showed greater improvement in the three evaluation dimensions by adding a cultural context prompt. Moreover, our mediation analyses revealed that cultural prompting improved perceived empathy in the therapeutic responses through increased cultural responsiveness. This indicates that cultural adaptation can improve the perceived empathy of therapeutic communication.

Our findings extend previous research on evaluating the efficacy of cultural prompting on cultural alignment by evaluating cultural value survey data[47], demonstrating the effectiveness of cultural prompting in improving cultural responsiveness in the context of a healthcare application. In a previous study that used CRQ for evaluating culturally adapted psychological intervention, a score of 3.5 on a scale of 1-5 was considered "average to good" and adequate to proceed to implement the culturally adapted psychological intervention[44]. In our findings, only the GPT model with cultural prompting achieved an average CRQ rating of 3.6 on a scale of 1-5, which would be considered an average good cultural adaptation by the standards of this previous work. While GPT-4o with cultural prompting emerged as the most preferred, GPT-4o without cultural prompting trended as the least preferred. This highlights the importance of adopting cultural prompting techniques for using LLMs such as GPT-4o for therapeutic intervention delivery. The effectiveness of this relatively straightforward cultural prompting technique suggests that meaningful improvement in cultural responsiveness and perceived empathy can be achieved without extensive system design, additional computing, or further human resources. While traditional cultural adaptation approaches require extensive resources pre-implementation[13,48], our results demonstrate that real-time adaptation can be achieved through appropriate prompt engineering.

The difference in the effectiveness of cultural prompting in GPT-4o and DeepSeek-V3 suggests that the effectiveness of cultural prompting is model-dependent, which aligns with previous findings in the cultural alignment of LLMs[17,19]. We observed that the responses generated by GPT-4o with cultural prompting had more tokens than the culturally prompted responses generated by DeepSeek-V3. That is, GPT-4o tends to prolong responses to integrate more cultural information. This difference in responsiveness may be attributed to architectural variations and differences in training protocols. Alternatively, the difference could be attributable to the fact that we optimized the therapy prompts based on GPT models[32], and did not further tailor and optimize them for DeepSeek.

The results of the mediation analysis provide compelling evidence that cultural responsiveness served as the mediator through which cultural prompting affects perceived empathy. This finding suggests that perceived empathy in AI-delivered therapeutic interventions for specific cultural populations can be effectively improved with cultural prompting. Our findings demonstrated that 40-50% of the improvement in perceived empathy was attributable to the improvement in cultural dimensions. This finding has important implications for LLM-based therapeutic interventions, suggesting that including cultural contexts in the prompt should not be supplementary but rather a core component to ensure the delivery of empathetic and effective interventions. Without addressing the cultural context, the LLM-based interventions may have limited effectiveness for culturally diverse populations.

Several limitations of the current study present opportunities for future work. First, the participants were predominantly highly educated (i.e., with a Bachelor's or more advanced degree), long-term immigrants (i.e., immigrated to the U.S. on average two decades ago). This limits the generalizability of findings, and future research should aim to include participants with more diverse educational backgrounds, immigration experiences, acculturation background, and diverse cultural-linguistic contexts (e.g., other languages and cultures) to enhance the generalizability of the findings. In this study, our approach to culturally adapted LLMs' responses was through system prompt augmentation, which relied heavily on pre-existing learned knowledge about a culture embedded within LLMs. Future studies could explore approaches incorporating external cultural knowledge, such as cultural sayings[21]. Future work should explore more sophisticated LLM-compatible cultural adaptation techniques, such as a well-designed workflow with multiple agents incorporating specific dimensions such as cultural values, cultural burdens, and communication styles. In addition, the current evaluation only examined single-turn conversation instances rather than a real-time, multi-turn therapeutic session. Future studies should explore the effects of techniques like cultural prompting in a multi-turn therapeutic session. A multi-turn therapeutic session will also allow us to evaluate more clinical outcomes and proxies for clinical outcomes, such as emotional state and therapeutic alliance.

**Conclusion**
Our study contributes to the literature on culturally responsive AI applications in healthcare by presenting a human-centered evaluation of the cultural responsiveness of LLM-generated responses in a therapeutic context. This study provides evidence that cultural prompting techniques can significantly improve the cultural responsiveness (competence and relevance) and perceived empathy of GPT-4o-generated therapeutic responses in supporting Chinese American family caregivers. Parallel evaluations of similar approaches on different languages and cultures are warranted. Furthermore, our findings revealed that cultural responsiveness mediated cultural prompting's improvements in perceived empathy, suggesting that cultural adaptation is not merely an additional feature but a core component for delivering empathetic AI-based health interventions to culturally and linguistically diverse populations. Our study also highlights that although the prompt-based technique offers a practical solution for cultural adaptation without requiring additional resources for model modification, the efficacy of cultural prompting is model-dependent. Future work should continue exploring additional techniques to improve cultural responsiveness and evaluate their effects on LLM-delivered therapy chatbots among diverse populations.

**Acknowledgments**
This work is supported, in part, by the University of Washington Population Health Initiative Pilot Grant, the National Institutes of Health (NIH) Agreement No.1OT2OD032581 and R21NR020634, and the Rita and Alex Hillman Foundation Emergent Innovation Program.